\documentclass[aps,prd,preprintnumbers,nofootinbib,twocolumn]{revtex4}

\usepackage{amsmath}
\usepackage{amsfonts}
\usepackage{amssymb}
\usepackage{graphicx}
\usepackage{color}

\newcommand{\be}{\begin{equation}}
\newcommand{\bea}{\begin{eqnarray}}
\newcommand{\eea}{\end{eqnarray}}
\newcommand{\ba}{\begin{array}}
\newcommand{\ea}{\end{array}}
\newcommand{\ee}{\end{equation}}
\newcommand{\bes}{\begin{equation*}}
\newcommand{\beas}{\begin{eqnarray*}}
\newcommand{\eeas}{\end{eqnarray*}}
\newcommand{\bas}{\begin{array*}}
\newcommand{\eas}{\end{array*}}
\newcommand{\ees}{\end{equation*}}
\newcommand{\nn}{\nonumber}

\def\eqb{\begin{eqnarray}}
\def\eqe{\end{eqnarray}}
\def\a{\alpha}
\def\h{\hbar}

\def\k{\kappa}
\def\d{\delta}

\def\it{\textit}
\begin{document}

\title{Discreteness of Space from GUP in a Weak Gravitational Field}
\author{Soumen Deb$^{1}$}\email[]{soumen.deb@uleth.ca}
\author{Saurya Das$^{1}$}\email[]{saurya.das@uleth.ca}
\author{Elias C. Vagenas$^{2}$}\email[]{elias.vagenas@ku.edu.kw}
\affiliation{$^1$~Theoretical Physics Group, Dept. of Physics and
Astronomy, University of Lethbridge, 4401 University Drive, Lethbridge, Alberta T1K 3M4, Canada}
\affiliation{$^2$~Theoretical Physics Group, Department of Physics, Kuwait University, P.O. Box 5969, Safat 13060, Kuwait}

\begin{abstract}
\par\noindent
Quantum gravity effects modify the Heisenberg's uncertainty principle to a generalized uncertainty principle (GUP). 
Earlier work showed that the GUP-induced corrections to the Schr\"{o}dinger equation, when applied to a non-relativistic 
particle in a one-dimensional box, led to the quantization of length. Similarly, corrections to the Klein-Gordon and the 
Dirac equations, gave rise to length, area and volume quantizations. These results suggest a fundamental granular structure 
of space. In this work, it is investigated how spacetime curvature and gravity might influence this discreteness of space. 
In particular, by adding a weak gravitational background field to the above three quantum equations, it is shown that 
quantization of lengths, areas and volumes continue to hold. However, it should be noted that the nature of 
this new quantization is quite complex and under proper limits, it reduces to cases without gravity. 
These results suggest that quantum gravity effects are universal.  
 \end{abstract}

\maketitle

\section{Introduction}
\par\noindent
During the last 70 years, much effort  has been devoted  towards the construction of a consistent 
theory of Quantum Gravity (QG). All approaches to QG start with an assumption about the structure of spacetime at scales 
that are extremely small, way beyond the current experimental advancement. 
\par\noindent
However, even if not direct, experimental evidence, e.g.  analogue gravity experiments \cite{Barcelo:2005fc}, 
suggests that gravity can show quantum effects. Therefore, since there is no direct experimental guidance,  
it is quite natural to try to develop a correct theory based on  conceptual restrictions. 
Like any other active research field,  what Quantum Gravity Phenomenology (QGP) ideally 
needs is a combination of theory and doable experiments \cite{AmelinoCamelia:2008qg}. 
\par\noindent
At the moment,  QGP can be thought of as a combination of all studies that might contribute 
to direct or indirect observable predictions \cite{Ali:2011fa,Pikovski:2011zk} and analog models \cite{Barcelo:2005fc}. 
These studies support the small and the large scale structure of spacetime consistent with String Theory, or any other  
approaches to QG. 
\par\noindent
The first step to identifying the relevant doable experiments for QGP research would be the identification of the 
working scale of this new field. This, known as the Planck scale, is first estimated from dimensional arguments. 
The Planck scale is uniquely defined by the fundamental constants, namely the speed of light $c$, 
the gravitational constant $G$, and the Planck constant $\hbar$, to provide the units of length, mass, and time 
\bea
\ell_{Pl}&=&\sqrt{\frac{\hbar G}{c^3}}\sim 10^{-35} m \nn~,\\
m_{Pl}&=&\sqrt{\frac{\hbar c}{G}}\sim 10^{-8} kg\nn~,\\
 t_{Pl}&=&\sqrt{\frac{\hbar G}{c^5}}\sim 10^{-44} s\nn ~.
\eea
\par\noindent
The smallness of this scale makes QG phenomenologists' job difficult, which is to test the Planck scale 
effects and extract useful information for further theoretical studies.
\par\noindent
Among the many mathematical results of String Theory there is one which is of particular interest and relevant to 
QGP. This is the modification of the Heisenberg uncertainty principle (HUP), which is well known  
as generalized uncertainty principle (GUP). 
In the context, mainly but not only, of String Theory, the suggested version of GUP is  \cite{Veneziano:1986zf, Gross:1987ar,Amati:1988tn,
Maggiore:1993rv,Konishi:1989wk,Garay:1994en,Scardigli:1999jh,Hossenfelder:2003jz,Bambi:2007ty}
\be 
\Delta x\geq \frac{\hbar}{\Delta p}+\alpha^{'}\frac{\Delta p}{\hbar}
\ee
where $\sqrt{\alpha^{'}}\approx 10^{-32}$ cm \cite{Witten:2001ib}.
\par\noindent
Recently, the  theories of Doubly Special Relativity (DSRs) were introduced principally to give a physical 
interpretation of the Planck length, i.e., $\ell_{Pl}$, in the structure of spacetime  \cite{AmelinoCamelia:2000mn}.  
In particular,  different values could be attributed to the Planck length by different observers.  
Thus, DSRs  avoid these violations of Lorentz invariance by considering the Planck length 
as an observer-independent scale. One of the consequences of DSRs  was  a 
similar modification of the position-momentum  commutation relation 
\cite{Magueijo:2004vv,Cortes:2004qn} which leads to a modification of the 
HUP as well. In this case, the  suggested form of the commutator is given by \cite{Ali:2011fa}
\bea
[x_i,p_j]&=&i\hbar \left(\delta_{i j}-\alpha\left(p\delta_{ij}+\frac{p_i p_j}{p}\right) \right.\nn\\
 &&\left. + \alpha^2(p^2\delta_{ij}+3p_i p_j)\right)
\label{commutator1}
\eea
where $p$ can be interpreted as the magnitude of $\vec{p}$ since 
$p^2=\sum\limits_{i=1}^3 p_i p_i$ and $\alpha=\frac{\alpha_0}{m_{Pl}c}=\frac{\alpha_0\ell_{Pl}}{\hbar}$.
\par\noindent
The suggested form of the commutator given in Eq. (\ref{commutator1}) is satisfied   by the modified operators 
\be
\begin{array}{ll}
x_i = x_{0i},\\
p_i  = p_{0i} (1-\alpha p_0+2\alpha^2 p_0^2), \hspace{0.5cm}i=1, 2, 3~.
\end{array}
\label{modifiedoperators1}
\ee
\par\noindent
Here, $x_{0i},~ p_{0i}$ satisfy the canonical commutation relations $[x_{0i},p_{0i}]=i\hbar\delta_{ij}$, 
 implying that $p_{0i}=-i\hbar\frac{\partial}{\partial x_{0i}}$ is the standard momentum (operator) at low energies and 
$p_i$ the  modified momentum at higher energies. Note that $p_0^2=\sum\limits_{i=1}^3p_{0i} p_{0i}$ \cite{Ali:2009zq}.\\
\par\noindent
The specific modification of the commutator (see Eq. (\ref{commutator1})), with the modified operators as given in 
Eq. (\ref{modifiedoperators1}), leads to a version of GUP which reads 
\cite{Kempf:1996fz,Kempf:1994su,Brau:1999uv}
\bea
\Delta x \Delta p &\geq & \frac{\hbar}{2}\left[1-2\alpha<p>+4\alpha^2<p^2>\right] \nn\\
 &\geq & \frac{\hbar}{2}\left[1+\left(\frac{\alpha}{\sqrt{<p^2>}}+4\alpha^2\right)\Delta p^2\right.\nn\\
 &&\left. + 4\alpha^2<p>^2-2\alpha\sqrt{<p>^2}\right]
 \eea
with the dimensionless parameter $\alpha_0$ generally considered to be of  order of unity.
\par\noindent
It is evident that QGP indicates an irremovable uncertainty in distance measurements \cite{AmelinoCamelia:2008qg}. 
In the framework of String Theory,  the modified commutation relations of position and momentum operators
 result in a version of GUP. A similar, but subtler, consequence of this version is that the apparently continuous-looking  space on a 
very fine scale is actually grainy. One can ask whether this is a sole influence of gravity or a fundamental structure of the 
spacetime. Now, if one admits the fact that classical gravity is a derived effect of curvature of spacetime caused by mass, 
then one can expect to find this discontinuity even in the regions of the universe far from a massive object. 
\par\noindent
The nature of this discreteness may or may not change when the spacetime is no more flat, 
namely it is a curved spacetime due to the presence of a gravitational field. In order to investigate this,
 we trap a particle in a box with a gravitational potential inside the box and see 
if gravity influences the discreteness shown in \cite{Ali:2009zq,Das:2010zf}.
\par\noindent
The outline of this work is as follows. In the next Section, we briefly review the  problem of a particle moving in a 
one-dimensional potential. Spacetime is flat but due to  GUP-effects, it effectively shows a discrete structure. 
In Sec. III, we investigate the discreteness of spacetime in the problem of a non-relativistic 
particle moving in a one-dimensional potential when gravity is present. Furthermore, we explore the discreteness of spacetime 
for the case of relativistic 0-spin and  1/2-spin particles  moving again  in a one-dimensional potential when gravity is 
 present. Finally, in Sec. V, we briefly present our results.
%
%
%
%
%
\section{Discreteness of space in flat spacetime}
%
%
%
\par\noindent
In this section, we briefly review the non-relativistic situation where a particle is trapped in a one-dimensional box  
and  one  finds the GUP-corrected Schr\"{o}dinger equation \cite{Ali:2009zq}. In particular, we consider  one 
of the standard examples in quantum mechanics, namely the  problem of a particle moving in a one-dimensional 
infinite potential well.  The well or the one dimensional box of length $L$ 
is defined by the potential $V(x)=0$ for $0\leq x \leq L$ and $\infty$ outside this box. 
The quantum mechanical equation governing such a particle is the Schr\"{o}dinger equation 
\be
H\psi=E\psi\nn
\ee
\par\noindent
 except for the fact that the position and momentum operators are now modified 
due to the GUP-effects. 
\par\noindent
Incorporating the GUP corrections, one can write the modified Schr\"{o}dinger equation as
\be
\frac{d^2}{dx^2}\psi+k_0^2\psi+2i\alpha\hbar\frac{d^3}{dx^3}\psi=0
\label{modifiedSchrodinger}
\ee
\par\noindent
where $k_0=\sqrt{2mE/\hbar^2}$.\\
At this point, it should be stressed that the $\alpha$-dependent term in the above equation is only important 
when energies are comparable to Planck energy and lengths are comparable to Planck length. 
The general solution of this equation is
\be
\psi=Ae^{ik_0^{'}x}+Be^{-ik_0{''}x}+Ce^{ix/2\alpha\hbar}~.\nn
\ee
\par\noindent
The first two terms along with the boundary conditions $V(x=0)=0=V(x=L)$ lead to the standard energy quantization. 
It is the new third $\alpha$-dependent term that gives rise to a new condition \cite{Ali:2009zq}
\be
\cos\left(\frac{L}{2\a\h}-\theta_C\right)=\cos(k_0L+\theta_C)=\cos(n\pi+\theta_C+\delta_0)\nonumber
\ee\
\par\noindent
which in turn implies that\footnote{As already mentioned, for brevity  the mathematical 
details have been omitted here. However, for the interested reader, the derivation of the quantization condition, 
i.e. Eq. (\ref{quantization1}), can be found in \cite{Ali:2009zq}. The whole analysis  
goes from Eq.(11) to Eq.(21) of  reference \cite{Ali:2009zq}.}
\be
\frac{L}{2\a\h}=\frac{L}{2\a_0\ell_{Pl}}=-n\pi+2q\pi\equiv p\pi
\label{quantization1}
\ee
\par\noindent
where $p\equiv2q\pm n$ is a natural number. 
The above expression shows that the length $L$ is  quantized. This result can be interpreted as the fact that, 
like the energy of the particle inside the box, the length of the box can assume only certain values. In particular, $L$ 
has to be in units of $\a_0\ell_{Pl}$. This indicates that the space, at least in a confined region and without the 
influence of gravity, is likely to be discrete.
\par\noindent
Further work has shown that this consequence of the  GUP effects can be extended to relativistic scenarios in one, 
two, and three dimensions \cite{Das:2010zf}. There are several reasons why one needs to investigate the relativistic cases. 
High energy particles are much more likely to probe the fabric of spacetime near the Planck scale, 
which means that they are necessarily relativistic or ultra-relativistic particles. In addition, the fact, that most elementary 
particles are fermions, leads us to investigate Dirac equation instead of the Schr\"{o}dinger equation. 
%
%
%
%
%
%

\section{Discreteness in curved spacetime}
\par\noindent
It has been proven that the GUP corrections imposed on a free particle  lead to the discreteness of space. 
Although the moving particle was kept in a box, no force field inside the box was assumed, 
i.e., the particle was free to move in  a flat spacetime.  If we wish to claim that the quantum gravity effects are 
universal then we should expect that the length quantization will also emerge in the presence of external forces. 
In other words, discreteness of space must hold whether or not there is an external field present. 
%
%
%
\subsection{Non-relativistic  case}
\par\noindent
The first step towards this generalization would be to consider gravity as the external force field inside the box, since
it is the weakest among the four fundamental forces as well as being universal. Additionally, as we have discussed before, 
our goal is to find how gravity determines the nature of discreteness. With a gravitational potential present inside  
the box, we ignore all but the first term of the Taylor expansion of this potential, which is  a linear term.  
This is reasonable because we are interested in the behavior of spacetime fabric near Planck scale and the gravitational 
potential changes very little over such small distances. Furthermore, in practice, we often use the 
gravitational potential energy approximated as $V(h)=mgh$ over a small vertical distance $h$ and, thus, the field strength reads
$E_h=-\frac{1}{m}\frac{\partial{d}V(h)}{\partial{d}h}=- g$. It is evident that this also justifies the previous claim 
of utilizing a linearized potential term.
%
%
%
%
%
%
%
%
\par\noindent
Let us now consider a one-dimensional box of length $L$ $(0< x < L)$ with a linear potential inside, 
which has the form
\be
V(x) = \left\{ \begin{array}{ll}
k x, & \textrm{if $0 < x < L$}\\
\infty, & \textrm{otherwise}\\
\end{array} \right.
\ee
with $k$ to be a parameter of unit $J/m$ and the smallness of $k$ is assumed. 
\par\noindent        
Without considering the GUP effects, the Schr\"{o}dinger equation governing the motion of a particle of mass $m$ 
inside this box is given by \cite{ll}
                
\be
\frac{d^2\psi_{\!_0}(x)}{dx^2}-\frac{2m}{\hbar^2}(kx-E)\psi_{\!_0}(x)=0~.
\label{airy}
\ee
\par\noindent
with $\psi_{\!_0}(x)=0$ when $x \le 0$ or $x \ge L$, since the potential outside the box becomes $\infty$.\\
The above equation, namely Eq.(\ref{airy}), is an Airy equation whose general solution reads \cite{ode}
\be
\psi_{\!_0}(x)=C_1 Ai\!\!\left[\frac{\frac{2m}{\hbar^2}(kx-E)}{(\frac{2m}{\hbar^2}k)^\frac{2}{3}}\right] + 
C_2  Bi\!\!\left[\frac{\frac{2m}{\hbar^2}(kx-E)}{(\frac{2m}{\hbar^2}k)^\frac{2}{3}}\right]
\label{solution0}
\ee
\par\noindent
where $Ai[u]$ and $Bi[u]$ are Airy functions of the first and second kind, respectively.
\par\noindent
We now use this wavefunction, i.e., $\psi_{\!_0}$, for solving the GUP-corrected  Schr\"{o}dinger equation.  
Utilizing the GUP-modified operators given in Eq. (\ref{modifiedoperators1}) in order to modify the Hamiltonian 
of the system under study, the GUP-corrected one-dimensional Schr\"{o}dinger equation for a non-relativistic particle 
moving in a box of length $L$ with a linear potential reads, cf. Eq. (\ref{modifiedSchrodinger}),
\be
\frac{d^2 \psi}{dx^2} +\frac{2m}{\hbar^2}(E-kx)\psi + 2i\alpha\hbar\frac{d^3 \psi}{dx^3}=0~.
\label{gupSchrodinger}
\ee
\par\noindent
It is  seen that the additional third term,  $2i\alpha\hbar\frac{d^3 \psi}{dx^3}$, which depends on 
the GUP parameter, i.e.,  $\a$, becomes significant at high energies (comparable to Planck energy), or, equivalently, at small 
lengths (comparable to Planck length). Therefore, we can consider a perturbative approach 
in order to solve Eq. (\ref{gupSchrodinger}).  A suitable trial solution can be of the form 
\bea
\psi_1 &=& \psi_{\!_0}(E+\epsilon\alpha, k, x) \nn\\
&=& \psi_{\!_0}(E, k, x)+\epsilon\alpha \frac{d}{dE}\psi_{\!_0}(E, k, x)
\label{solution1}
\eea
\par\noindent 
where the form of $\psi_0$ is given by Eq. (\ref{solution0}), and $\epsilon$ is a coefficient that will be 
determined later. 
\par\noindent
 Skipping  intermediate mathematical steps, the general solution of the GUP-corrected 
Schr\"{o}dinger equation is given by
\bea
\psi(x)
&=&\frac{A}{\sqrt\pi}\left[\xi^{-1/4} \sin\left(\frac{2}{3}\xi^\frac{3}{2}+\frac{\pi}{4}\right)+\right.\nn\\
&&\left. \left(\frac{2m}{\hbar^2}\right)^{1/3}k^{-2/3}\epsilon\alpha\left(-\frac{1}{4}\xi^{-5/4}\sin\left(\frac{2}{3}\xi^{3/2}+\frac{\pi}{4}\right)\right.\right.\nn\\
&&\left.\left.+\xi^{1/4}\cos\left(\frac{2}{3}\xi^{3/2}+\frac{\pi}{4}\right)\right)\right]+\nn\\
&&\frac{B}{\sqrt\pi}\left[\xi^{-1/4} \cos\left(\frac{2}{3}\xi^\frac{3}{2}+\frac{\pi}{4}\right)+\right.\nn\\
&&\left.\left(\frac{2m}{\hbar^2}\right)^{1/3}k^{-2/3}\epsilon\alpha\left(-\xi^{1/4}\sin\left(\frac{2}{3}\xi^{3/2}+\frac{\pi}{4}\right) \right.\right.\nn\\
&&\left.\left. -\frac{1}{4}\xi^{-5/4}\cos\left(\frac{2}{3}\xi^{3/2}+\frac{\pi}{4}\right)\right)\right]+Ce^{ix/2\hbar\alpha}
\label{solution1general}
\eea
with 
\bea
\xi  &=& \left(\frac{2m}{\hbar^2}\right)^{\frac{1}{3}}k^{-\frac{2}{3}}\left(E-kx  \right)\nn\\
\epsilon &=&
\left[ 
\left(2 i \hbar \right) \frac{3}{4} \left( \frac{2m}{\hbar^2}\right)^{\frac{11}{12}} 
k^{\frac{7}{6}} E^{-\frac{1}{4}} \right. \nn\\
&\times& 
\left( 
C_{1} \sin \left( \xi_{0}+\frac{\pi}{4}\right) - C_{2} \cos \left( \xi_{0}+\frac{\pi}{4}\right) 
\right) \nn\\
&+&
\alpha \left( 2 i \hbar \right) \left( \frac{2m}{\hbar^2}\right)^{\frac{17}{12}} 
k^{\frac{1}{6}} E^{\frac{5}{4}} \nn\\
&\times& \left. 
 \left( C_{2} \sin \left( \xi_{0}+\frac{\pi}{4}\right) - C_{1} \cos \left( \xi_{0}+\frac{\pi}{4}
\right) 
\right)
 \right]\nn\\
&\div&
\left[ 
\left(\frac{2m}{\hbar^2}\right)^{\frac{11}{12}} k^{\frac{1}{6}} E^{-\frac{1}{4}} \right.\nn\\
&\times&\left.
 \left( C_{1} \sin \left( \xi_{0}+\frac{\pi}{4}\right) - C_{2} \cos \left( \xi_{0}+\frac{\pi}{4}\right) 
\right)  \right]\nn\\
\xi_{0}  &=& \frac{2}{3}\left(
 \left(\frac{2m}{\hbar^2}\right)^{\frac{1}{3}}k^{-\frac{2}{3}} E 
\right)^{\frac{3}{2}}\nn
\eea
and $A$, $B$, $C$ are constants. 
In particular, we can absorb the phase of $A$ in $\psi$, such that $A$ can be treated as a real  constant while $B$ can be 
written as $B=|B| e^{i\theta_B}$. Furthermore, $C$ is such a constant that its magnitude $|C|$ 
becomes zero in the limit $\alpha \rightarrow 0$, since the last term must vanish in this limit.
\par\noindent
Next, by imposing  the boundary conditions $\psi(x=0)=0$ and $\psi(x=L)=0$, we arrive 
at the following condition on the length of the box
\bea
\cos(L/2\hbar\alpha) 
\!&=&\!\left(1-\frac{kL}{E}\right)^{-1/4}\times\nn\\ 
\!&&\!\left[\!A_*\sin\left(\frac{2}{3}\sqrt{\frac{2m}{\hbar^2}}\frac{(E-kL)^{3/2}}{k}+\frac{\pi}{4}\right)\!+\!\right.\nn\\
\!&&\!\left. B_*\cos\left(\frac{2}{3}\sqrt{\frac{2m}{\hbar^2}}\frac{(E-kL)^{3/2}}{k}+\frac{\pi}{4}\right)\!\!\right]
\label{condition1}
\eea
\par\noindent            
where $A_*$ and $B_*$ are constants that depend on $A$, $B$, $k$, and $E$. 
\par\noindent
It can be shown that in the limit $k\rightarrow 0$ the wavefunction given by Eq. (\ref{solution1general}) 
 becomes the solution of Schr\"{o}dinger equation for an infinite potential well. Thus, taking the limit  $k\rightarrow 0$
the RHS of Eq. (\ref{condition1}) reads
\be
B_{1} \cos \left( \kappa L_{0} \right) - A_{1} \sin \left(  \kappa L_{0} \right)
\ee
where $L_{0}$ is the length of the box in flat spacetime, $\kappa=\sqrt{\frac{2mE}{\hbar^{2}}}$, and 
\bea
A_{1}\!\!&=&\!\! H_{1}\!\left(  A_{*} \cos\left( \frac{H_{2}}{k}+\frac{\pi}{4}\right) -  
B_{*} \sin\left( \frac{H_{2}}{k}+\frac{\pi}{4}\right)\!\!\right)\hspace{2ex}  \\
B_{1}\!\!&=&\!\! H_{1}\!\left(  A_{*} \sin\left( \frac{H_{2}}{k}+\frac{\pi}{4}\right) + 
B_{*} \cos\left( \frac{H_{2}}{k}+\frac{\pi}{4}\right)\!\!\right)\hspace{2ex} 
\eea
with
\bea
H_{1} &=&\frac{1}{\sqrt{\pi}}\left[ \left(\frac{2m}{\hbar^{2}}\right)^{\frac{1}{3}} \frac{\left(E-kx\right)}
{k^{\frac{2}{3}}} \right]^{-\frac{1}{4}} \\
H_{2} &=&\frac{2}{3} \left(\frac{2m}{\hbar^{2}}\right)^{\frac{1}{2}} E^{\frac{3}{2}}~.
\eea
\par\noindent
Without loss of generality, we let $A_{1} = \sin \theta$ and $B_{1} = \cos \theta$ for an arbitrary $\theta$; thus, 
Eq. (\ref{condition1}) becomes
\bea
\cos(L_{0}/2\hbar\alpha)  & = & \cos \theta \cos\left( \kappa L _{0}\right) - \sin\theta \sin\left( \kappa L_{0} \right)\\
&=& \cos \left(   \kappa L_{0} +\theta \right)~.
\eea
\par\noindent
According to the analysis in \cite{Ali:2009zq}, the above equation implies that $\frac{L_0}{2\hbar\alpha}=p\pi$, 
$p \in \mathbb{N}$. Since $L$ is the perturbation of $L_{0}$,  Eq. (\ref{condition1}) yields
\be
\frac{L}{2\hbar\alpha}=f(k)p_1\pi+p\pi
\label{quantization2}
\ee
\par\noindent
where $p_{1} \in \mathbb{N}$ and for each $p$ there is a finite set of values of 
$p_1 \in \mathbb{N}$. Moreover, since the first term on the RHS of Eq. (\ref{quantization2}) is  a small 
perturbative term,  the number  of $p_1$ values, for each 
$p$, depends on the smallness of  function $f(k)$. 
\par\noindent 
As in the case of flat spacetime, we have arrived at a length quantization condition. Moreover, we have a fine 
structure (splitting) of the length quantization due to the presence of  gravity (see Fig. 1). 
This is similar to the energy quantization of the hydrogen atom, 
in presence of an external electromagnetic field.
\hspace{1cm}\begin{figure}
\includegraphics[scale=.5]{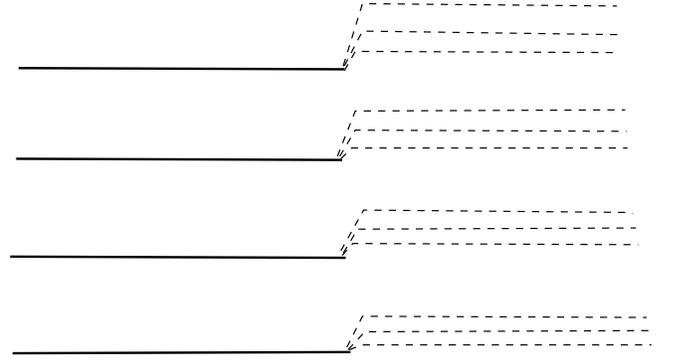}
\caption{Comparison between $L_0$ (solid lines) which is the quantized length with GUP corrections
in flat spacetime and $L$ (dotted lines) which is the quantized length  with GUP corrections in curved spacetime.}
%
%
%
\end{figure}
%
%
%
\subsection{Relativistic Case}
\par\noindent
The small-scale structure of spacetime should not depend on the use of relativistic or non-relativistic test 
particles. However,  particles with speeds comparable to the speed of light should be treated 
relativistically, and the fundamental spacetime structure should be reexamined. 
In this subsection, we take a closer look at the relativistic 
equivalent of the Schr\"{o}dinger equation, i.e., Klein-Gordon equation and, in particular, the modification 
induced by GUP. First,  we will derive the GUP-version of the Klein-Gordon equation with a linear potential 
and then we will try to solve it in order to obtain possible length quantization. Notwithstanding its relative simplicity, 
Klein-Gordon equation has mathematical difficulties, especially when it comes to dimensions higher than one. 
For this reason, it is much easier to implement the more versatile Dirac equation. Therefore, we will also 
solve the Dirac equation in order to find a similar length quantization as in \cite{Das:2010zf}.
%
%
%
\subsubsection{Klein-Gordon Equation}
\par\noindent
The Klein-Gordon  equation with no force field is given by \cite{ad}
\be
(\hbar^2\Box+m^2c^2)\psi=0
\ee
\par\noindent
where $\Box=\frac{1}{c^2}\frac{\partial^2}{\partial t^2 }-\nabla^2$ and  
$\vec{\nabla}=\frac{\partial}{\partial x}\hat{i}+\frac{\partial}{\partial y}\hat{j}+\frac{\partial}{\partial z}\hat{k}$.
\par\noindent
Next, we take into account a  gravitational force field  by utilizing a linearized potential. 
In this case, the GUP-corrected Klein-Gordon equation in one dimension  reads
\be
\!\frac{d^2\psi}{dx^2}+\frac{1}{\hbar^2c^2}\left(E^2-m^2c^4-2Ek x\right)\psi 
+2i\alpha\hbar\frac{d^3\psi}{dx^3}=0~.\!
\label{gupKG}
\ee
\par\noindent          
Comparing Eq. (\ref{gupKG}) with Eq. (\ref{gupSchrodinger}), i.e.,
\be
\frac{d^2\psi}{dx^2}+\frac{2m}{\hbar^2}(E-kx)\psi  + 2i\alpha\hbar\frac{d^3\psi}{dx^3}=0~, \nn
\ee
\par\noindent
and by making the following    ``transformations"   
\bea
&&\frac{2m}{\hbar^2}E\rightarrow\frac{1}{\hbar^2 c^2}(E^2-m^2 c^4) \nn\\
&&\frac{2Ek}{\hbar^2 c^2}\rightarrow \frac{2mk}{\hbar^2}\nn
\eea
\par\noindent
we arrive at a length quantization similar to the one  given by Eq. (\ref{quantization2}).
%
%
%
\subsubsection{Dirac Equation in one dimension}
\par\noindent
The three-dimensional version of Klein-Gordon equation suffers from the non-locality of the differential operators. 
In particular, the term $p^2$, when GUP is considered, becomes 
\be
p^2 = p_0^2-2\a p_0^3=-\h^2\nabla^2+2i\a\h^3\nabla^3  \nn
\ee
\par\noindent
and, thus,  the second term reads
\be
2i\a\h^3\left(\frac{\partial^2}{\partial x^2}+\frac{\partial^2}{\partial y^2}+\frac{\partial^2}
{\partial z^2}\right)^{3/2}\nn~. 
\ee
\par\noindent
We can deal with this term which is a non-local one using fractional calculus \cite{bolo}, but a much 
simpler approach would be to employ the Dirac equation.
\par\noindent
The free-particle Dirac equation is given by \cite{Dirac:1928hu}
\be
i \frac{\partial \Psi}{\partial t}=\left(\beta mc^2+c \vec{\alpha}\cdot\vec{P}\right)\Psi
\ee
\par\noindent   
with
\be 
\beta \equiv \gamma^0 = 
 \left( \begin{array}{cc}
{\bf I}_2 & 0 \\
0 &  -{\bf I}_2 
\end{array} \right)
\ee
\par\noindent          
and
\be
\alpha^i \equiv\gamma^0\gamma^i= 
\left( \begin{array}{cc}
{\bf I}_2  & 0 \\
0  & -{\bf I}_2 
\end{array} \right)
\left(\begin{array}{cc}
0 & \sigma_i\\
-\sigma_i & 0
\end{array}\right)\nn\\
= \left(\begin{array}{cc}
0 & \sigma_i\\
\sigma_i & 0 
\end{array}\right)
\ee
\par\noindent
where $\sigma_i$, with  i=x, y, z for the 3 spatial dimensions, are the Pauli spin matrices. These matrices are given by \cite{cohen}
\bea
&&  
 \sigma_x= 
\left(\begin{array}{cc}
0 & 1\\
1 & 0\end{array}\right),\hspace{.3cm}
\sigma_y= 
\left(\begin{array}{cc}
0 & -i\\
i & 0\end{array}\right),\hspace{.3cm}\nn\\
&&\sigma_z=\left(\begin{array}{cc}
1 & 0\\
0 & -1\end{array}\right)
\eea
\par\noindent
Here $\beta mc^2+c \vec{\alpha}\cdot\vec{P}$ is the Dirac Hamiltonian with no force field to be present. 
It should be noted that $\vec{\a}$ is distinct from the GUP  parameter, i.e.,  $\alpha$. 
\par\noindent
At this point, we take into account a  gravitational force field  by utilizing a potential term in the form $V(\vec{r})$. 
In this case, the GUP-corrected Dirac equation   reads
\bea
i\frac{\partial \Psi}{\partial t}=\left(\beta mc^2+c\vec{\alpha}\cdot\vec{P}+V(\vec{r}) {\bf  I_4}\right)\Psi~.
\eea
\par\noindent
Specifically, for the case of one spatial dimension, say $z$, the GUP-corrected Dirac equation reads
\be
\left(-ic\hbar {\bf \alpha}_z\frac{d}{dz}+c\alpha\hbar^2\frac{d^2}{dz^2}+\beta mc^2+kz {\bf I}_4\right)\psi(z)
=E\psi(z)~.\nn
\ee
\par\noindent
This equation represents a relativistic particle in a one-dimensional box with a potential of the form $kz$ inside.
\par\noindent
The four linearly independent solutions to this equation are given by
\bea
\psi_1&=&N_1\left(1-\frac{4ik\a\kappa z}{c/z+2i\a\kappa \left(c(1-2\a\kappa\h^2)-2E\right)}\right) \times\nn\\
&&e^{i\kappa z}
\left(\begin{array}{cc}
\chi\\
r\sigma_z\chi\end{array}\right)\nn\\
\psi_2 &=& N_2 e^{iz/\a\hbar}\left(\begin{array}{cc}
\chi\\
\sigma_z\chi \end{array}\right)
\eea
\par\noindent
with  $\chi$ to be a normalized spinor that satisfies the relation $\chi^{\dagger}\chi =I$.
\par\noindent        
Imposing the boundary conditions directly here, we end up having the so-called Klein paradox. 
In order to avoid this,  we will resort to the MIT bag model of quark confinement \cite{Chodos:1974dm}.  
Imposing the MIT bag boundary conditions and omitting some straightforward steps,  
the condition on the length of the box is given by
\bea
\frac{L}{\a\h} &=& arg\left[\frac{\rho_1(ir-1)\left(e^{i(\delta-\kappa L)-e^{i\left(\kappa L-tan^{-1}
\left(\frac{2r}{r^2-1}\right)\right)}}\right)}{F^{'}}\right]\nn\\
&& -\frac{\pi}{4}+2n\pi,~n \in \mathbb{N},
\eea
\par\noindent
where $\kappa=\kappa_0+\alpha\hbar \kappa_0^2$ with  $\kappa_0$ to be the wavenumber that satisfies the relation 
$E^2 = (\hbar\kappa_0)^2  +(mc^2)^2$. Additionally,  $r$, $\delta$, $\rho_1$, and $F'$ are defined as
\bea
r&=&\frac{\hbar \kappa_0 c}{E+mc^2}\nn\\
\delta&=&\tan^{-1}\left(\frac{2r}{r^{2} - 1}\right)\nn\\
\rho_1 &=&\left(1-\frac{4ik\a\kappa z}{c/z+2i\a\kappa\left(c(1-2\a\kappa\h^2)-2E\right)}\right) \nn\\
F' &=& \sqrt{2}F\nn
\eea
\par\noindent
with $F\sim \alpha^s$ and $s>0$.
%
%
%
\subsubsection{Dirac Equation in Three Dimensions}
\par\noindent
In the most general case, let us consider a box defined by $0\leq x_i \leq L_i, ~ i=1\dots d$, 
$d$ being the dimension of the box, 
i.e., $d=1, 2$, or $3$. That is, this box can be one, two, or three-dimensional. The box has a linearized potential inside, as before. 
Without loss of generality, we orient the box such that the direction in which the potential changes is our $x$-direction. 
The Dirac Hamiltonian with the linear potential term can now be written as
\bea
H&=&c \vec{\alpha} \cdot \vec{p}+\beta mc^2 +V(\vec{r})I\nonumber\\
&=& c\left(\a_x p_x+\a_y p_y+\a_z p_z\right)+\beta mc^2 + kxI\nonumber\\
&=& c\vec{\a}\cdot\vec{p_0}-c\a\left(\vec{\a}\cdot\vec{p_0}\right)\left(\vec{\a}\cdot\vec{p_0}\right)+\beta mc^2 + 
kxI~.\nn
\eea
\par\noindent      
Note that we employed the GUP-corrected momenta, i.e., 
$p_i=p_{0i}(1-\a p_0),~i=1,..,3$, where $p_{0i}=-i\h\frac{d}{dx_i}$, and followed Dirac prescription, i.e., 
we replaced $p_{0}$ by $\vec{\a}\cdot\vec{p_0}$.
\par\noindent
The wavefunction inside the box turns out to be
\be
\psi=\left(
\begin{array}{cc}\left[\prod\limits_{i=1}^d\left(\rho_1^{\d_{i1}}e^{i\k_i x_i}+\rho_2^{\d_{i1}}e^{-i(\k_i x_i-\d_i)}\right)\right.\nn\\
\left. +Fe^{i\frac{\hat{q}.\vec{r}}{\a\h}}\right]\chi\\
\sum\limits_{j=1}^{d}\left[\prod\limits_{i=1}^d\left(\rho_1^{\d_{i1}}e^{i\k_ix_i}+\right.\right.\nn\\
\left.\left.(-1)^{\d_{ij}}\rho_2^{\d_{i1}}e^{-i(\k_ix_i-\d_i)}\right)r\hat{\k_j}\right.\nn\\
\left.+Fe^{i\frac{\hat{q}.\vec{r}}{\a\h}}\hat{q_j}\right]\sigma_j\chi\end{array}\right)\nn\\
\ee
\par\noindent
where $\d_{ij}$ is the usual Kronecker delta, $\hat{q}$ is an arbitrary unit vector, and $\delta_l$ is given by
\be
\delta_{l}=\kappa_{l}L_{l}=\tan^{-1}\left( \frac{2r\hat{\kappa}_l}{r^{2}\hat{\kappa}^{2}_{l}-1} \right) 
+ \cal{O}(\ln\alpha)\nn
\ee
\par\noindent
with $\hat{\kappa}_l$ to be the $l$ component of the unit vector of the wave vector $\vec{\kappa}$ with components $\kappa_l$.
\par\noindent
Moreover, $\rho_1$ and $\rho_2$ are defined as 
\bea
\rho_1 &=&\left(1-\frac{4ik\a\kappa_{1} x}{c/x+2i\a\kappa_{1}\left(c(1-2\a\kappa_{1}\h^2)-2E\right)}\right) \nn\\
\rho_2 &=&\left(1+\frac{4ik\a\kappa_{1} x}{c/x-2i\a\kappa_{1}\left(c(1+2\a\kappa_{1}\h^2)-2E\right)}\right)~. \nn
\eea
The number of terms in the first row  is $2^d + 1$ and that in the second row  is $(2^d+1)\times d$.
\par\noindent
Using the MIT bag model again, we obtain conditions on the dimensions of the box. In this case, these 
conditions are not symmetrical unlike the case in flat spacetime. 
Along $x$-direction, the length quantization has the 
following form
\bea
 \frac{\hat{q}_1L_1}{\a\h}&=&\frac{\hat{q}_1 L_1}{\a_0\ell_{Pl}}\nn\\
&=&-\theta_1+ arg\left(\frac{\rho_1(ir\hat{\k}_1-1)-\rho_2(ir\hat{\k}_1+1)e^{i\d_1}}{F^{'}}f_{\bar{1}}\right)\nn\\
&&+2n_1\pi, ~n_1\in \mathbb{N}
\eea
\par\noindent
with  $f_{\bar{l}}(x_i, \kappa_i, \delta_i)=\prod\limits^{d}_{i=1 ( i\neq l)}
\left( e^{i\kappa_{i}x_{i}}+e^{-i(\kappa_{i}x_{i}-\delta_{i})}\right)$.
Along $y$ and $z$ directions, the quantization conditions are identical
\be
\frac{\hat{q_l}L_l}{\a\h}=\frac{\hat{q_l}L_l}{\a_0\ell_{Pl}}=-2\theta_l+2n_l\pi 
\ee
\par\noindent
with $n_l\in  \mathbb{N}$ and $\theta_{l}=\tan^{-1}(\hat{q}_{l})$.
\par\noindent          
This is also consistent with the fact that the potential inside the box increases linearly along $x$-direction and remains 
zero along $y$ and $z$ directions.
\par\noindent                
To obtain the  area and volume quantizations, we simply multiply the above conditions
\bea
A_N &=&\prod\limits_{l=1}^{N}\frac{\hat{q}_l L_l}{\a_0\ell_{Pl}}=\prod\limits_{l=2}^{N}\left(2n_l\pi-2\theta_l\right)
\left(2n_1\pi-\theta_1 \right.\nn\\
&+&\!\!\!\left.arg\!\left(\frac{\rho_1(ir\hat{\k}_1 -1)-\rho_2(ir\hat{\k}_1 +1)e^{i\d_1}}{F^{'}}f_{\bar{1}}\right)\!\right)
\eea
\par\noindent              
with $n_l\in \mathbb{N}$, and where $N=2$ and $N=3$ represent the area and volume quantization, respectively. \\
%
%
%
\section{Conclusions}
\par\noindent
In this work, we have shown that if we trap a particle in a one-dimensional box of size $L$, 
include a gravitational potential inside the box and then try to measure the length of the box, the length $L$ will appear as 
a quantized quantity in units of $\a_0 \ell_{Pl}$ where $\ell_{Pl}$ is the Planck length. This result can be interpreted as 
the discreteness of space near the Planck scale  holding for curved spacetime as it holds for  flat spacetime,  
as shown in previous works  \cite{Ali:2009zq,Das:2010zf}.
\par\noindent
For the gravitational potential, we have used the first term of a Taylor series to describe it  as 
a linearized potential. This is reasonable because we are interested in the behavior of spacetime fabric near Planck 
scale and the gravitational potential changes at a very slow rate over such small distances.
\par\noindent
We have implemented our method for a non-relativistic particle in curved spacetime and for a relativistic one. 
In the latter case, the GUP-corrected Klein-Gordon equation in one dimension has been solved as well as 
the GUP-corrected Dirac equation in one, two and three dimensions. As already mentioned, in all cases the length of the box 
appears as a quantized quantity in units of $\a_0 \ell_{Pl}$. The presence of lengths that are proportional to the 
Planck length in all cases strengthens the claim of the existence of a minimum measurable length. Furthermore, 
in two and three dimensions, the area and volume quantizations were also obtained. 
\par\noindent
Extension of the method employed in this work for arbitrary curved spacetime would be quite interesting.  In particular, 
it is expected that subsequent terms in the Taylor series would give rise to a more general curved spacetime. 
Hence, an arbitrary form of the gravitational potential could be analyzed following the same approach. 
This would still assume a fixed classical background.  A complete theory of quantum gravity, once  formulated, 
should be able to address the issues discussed here, with background spacetime which may be fluctuating. 
In this case, we hope that the results derived in this work would continue to hold, at least approximately, 
and almost exactly in the limit when such fluctuations can be ignored. 
%
%
%
%
%
%
%
\par\noindent            
Finally, one may be interested in delving into the possible connection between the non-relativistic particle moving 
in a box inside  which a linear potential is present, and the hydrogen atom. In both systems, a fine structure (splitting) shows up. 
In particular, for the first system it is the fine structure of the length quantization, while for second system it is the fine structure 
of the  energy quantization. This apparent coincidence suggests further  investigation of the  
discreteness of spacetime. In addition, although the original HUP is restricted to position-momentum 
commutation while the time-energy uncertainty principle has been merely  thought of as a statistical measure of variance, 
a more generalized idea of GUP-corrected commutation relation involving 
4-momentum might give rise to discontinuity of time.
%
%
%
%
%
\section{Acknowledgments}
\par\noindent
We would like to thank the referee for constructive comments. 
This work  is supported in part by the Natural Sciences and Engineering Research Council of Canada.
%
%
%

\end{document}